\documentstyle[12pt]{article}
\begin{document}
\baselineskip=24pt
\def\s{\section}
\def\ss{\subsection}
\def\sss{\subsubsection}
\def\ni{\noindent}
\def\hf{\hfill\break}
\def\i{\item}
\def\prl{Phys. Rev. Lett.}
\def\el{Europhys. Lett.}
\def\jp{J. Phys.}
\def\cs{Curr. Sci.}
\def\zp{Z. Phys.}
\def\pr{Phys. Rev.}
\def\be{\begin{equation}}
\def\ee{\end{equation}}
\def\ni{\noindent}
\date{}
\begin{center}
{\large \bf Statistics of Mesoscopic Fluctuations of Quantum Capacitance}
\vspace{0.80 cm}

N. Kumar$^*$ and A.M.Jayannavar$^{**}$\\
International Centre for Theoretical Physics\\
34100 Trieste, Italy\\

\end{center}

\ni{\bf Abstract}

The Thouless formula \(G = (e^2/h)(E_c/\Delta)\) for the two-probe dc
conductance $G$ of a d-dimensional mesoscopic cube is re-analysed to
relate its quantum capacitance $C_Q$ to the reciprocal of the level
spacing $\Delta$. To this end, the escape time-scale $\tau$ occurring in
the Thouless correlation energy \(E_c = \hbar/\tau\) is interpreted as the
{\em time constant} \(\tau = RC_Q\) with $RG \equiv$ 1, giving at once
\(C_Q = (e^2/2\pi \Delta)\). Thus, the statistics of the quantum capacitance is
directly related to that of the level spacing, which is well known from
the Random Matrix Theory for all the three universality classes of
statistical ensembles. The basic questions of how intrinsic this quantum capacitance
can arise purely quantum-resistively, and of its observability {\em
vis-a-vis} the external geometric capacitance that combines with it in series, are discussed.

\noindent
$\_\_\_\_\_\_\_\_\_\_\_\_\_$\\
$^*$Permanent address: Raman Research Institute, Bangalore 560080, India\\
$**$Permanent address: Institute of Physics, Bhubaneswar 751005, India\\

The non-self-averaging nature of quantum conductance, the transmittance
and the energy-level spacing of a mesoscopic system, due ultimately to
quantum coherence and the interference, has motivated extensive studies of
the statistics of their sample-to-sample fluctuations.$^1$  For a given
sample, however, these reproducible fluctuations manifest as random
functions of some externally controlled parameters such as the Fermi
energy, or the Aharonov-Bohm magnetic flux, etc. Microscopic analysis of the
statistics of these fluctuations in disordered electronic structures
having diffusive transport has been based on the impurity-diagrammatic
perturbative technique, while the macroscopic approaches generally appealed
to the random matrix theory.$^1$  Fluctuations of the wave transmission
coefficient and of the eigenvalue spacing for ballistic microstructures
(classically chaotic cavities) have been treated through the semiclassical
and the  random matrix approaches.$^{2,3}$  Very recently, quantum
capacitance$^{4,5}$ has been added to the growing list of these
non-selfaveraging quantities, where the authors$^6$ have calculated the
statistics of the mesoscopic capacitance fluctuations for the case of a
chaotic cavity coupled capacitatively to a backgate. This has involved
calculating the Wigner delay time which is given by the energy derivative
of the phase associated with the scattering matrix. The latter in 
turn was
obtained from a global random matrix theory for all the three universality
(symmetry) classes of the statistical ensembles. It may be noted in
passing that, for the case of a single channel, the distribution of the
time delay for a disordered mesoscopic system was obtained earlier using
the invariant imbedding approach.$^7$ In the present work we have addressed
the problem of calculating the statistics of the quantum capacitance of a
mesoscopic disordered conductor {\em directly} in terms of a generally
valid argument predicated on the Thouless expression for the conductance
\(G = (e^2/h)(E_c/\Delta)\) for a d-dimensional cube. Here $\Delta$ is the
energy level spacing at the Fermi level and \(E_c = \hbar/\tau\) the
Thouless (correlation) energy, where $\tau$ is the time scale of escape
out of the sample. Results for the {\em RC-time constant} for
the ballistic, the diffusive as well as for the possibly anomalous
diffusive regime at the mobility edge are reported. We also address the
physical question of how a reactive element such as the capacitance is
determined resistively through the Thouless conductance expression.

Starting from the Thouless expression for the conductance \(G =
(e^2/h)(E_c/\Delta)\), and recalling that \(E_c = \hbar/\tau\), where
\(\tau \equiv RC_Q\) is to be interpreted as the {\em RC-time constant}
with $C_Q$ as the effective (quantum) capacitance, and \(R(\equiv 1/G)\)
the series quantum resistance, we get at once \(C_Q = (e^2/2\pi
\Delta)\). Here the level spacing $\Delta$ scales as $L^{-d}$ for a
d-dimensional sample. Thus, we notice immediately that the quantum
capacitance $C_Q$ is a volume effect a (proportional to $L^{-3}$ in three
dimensions), quite unlike the classical (geometrical) capacitance that
scales as $L^{-1}$ in three dimensions.

Now the problem of finding the statistics of $C_Q$ reduces to that of
finding the level-spacing distribution. Defining the dimensionless
capacitance \(x = C_Q/\bar{C}_Q\), with \(\bar{C}_Q\) as the ensemble
averaged capacitance, we can at once write down the probability
distribution \(P_\beta(x)\) for the three universality classes, labelled by
the symmetry parameter $\beta$ = 1, 2 and 4 corresponding, respectively,
to the Gaussian Orthogonal Ensemble (GOE), the Gaussian Unitary Ensemble
(GUE) and the Gaussian Symplectic Ensemble (GSE), as

\begin{equation}
P_1(x) = \frac{1}{2\pi x^3} exp[-(1/\pi x^2)]\,\,,
\end{equation}
with the average capacitance \(\bar{C}_Q = (e^2/4<\Delta>)\);
\be
P_2(x) = \frac{\pi}{2x^4} exp[-(\pi / 4x^2)]\,\,,
\ee
with the average capacitance \(\bar{C}_Q = (2e^2/\pi^2<\Delta>)\);
and
\be
P_4(x) = (81\pi^2/128x^6) exp[-(9\pi/16x^2)],
\ee
with the average capacitance \(\bar{C}_Q = (16e^2/9\pi^2<\Delta>)\).
In all the three cases above, the quantum capacitance distribution has a
single peak and a long tail. In fact the variance diverges (weakly, i.e.,
logarithmically) for the GOE case. For a mesoscopic sample of interest, we
have typically $\Delta \sim$ 10 mK, and this gives $C_Q \sim$ 20 fF.

The fact that the quantum capacitance is determined resistively via the
Thouless formula has several consequences for the RC relaxation time of
the mesoscopic sample, coupled capacitatively to a backgate. This can be
seen as follows for the various transport regimes of interest.
\vspace{0.2cm}

\ni {\sl Ballistic regime}\\
Here the Thouless energy \(E_c \equiv (\hbar^2/2m) \int \mid
\bigtriangledown\psi\mid^2 d^dx\), (being the extra kinetic energy
associated with the phase-twisting of the wavefunction $\psi$ from the
periodic to the anti-periodic boundary condition) scales as $E_c \sim
L^{-1}$. Recalling that $\Delta \sim L^{-d}$, we get at once $G \sim
L^{d-1}$ (the two-probe Sharvin conductance of an ideal metal with
$L^{d-1}$ transverse channels). The corresponding RC-time constant $\tau$
scales as $\tau \sim L$.
\vspace{0.2cm}

\ni {\sl Diffusive limit}\\
Arguing as above, we have $E_c \sim 1/L_{arc}$, where $L_{arc}$ is the arc-
length of a random walk with the end-to-end distance $\sim L$ (sample
size). Thus, $L_{arc} \sim L^2$ giving RC-time constant $\tau \sim L^2$.
\vspace{0.3cm}

\ni {\sl Anomalous diffusion}\\
At and near the mobility edge when the sample size $L \ll \xi$ (the
coherence length), we can expect the diffusion to become anomalous in that
\(L_{arc} \sim L^{2\eta}\) with $\eta \neq 1$.  Reasoning as above, this
would give RC time constant $\tau \sim L^{2\eta}$.

\noindent
The above clearly shows the characteristically different scaling behaviour
of $\tau$ for the three different transport regimes. 

We now return to the
question of how a reactive element such as the quantum capacitance can
arise concomitantly with the quantum resistance. The whole point is that
on the one hand the quantum capacitance (unlike its classical geometrical
counterpart) involves the energy cost of promoting the piled-up electrons
across the energy-level spacing, while on the other the latter also
represents the typical energy mismatch (obstructing the electron
diffusion) which is involved in determining the conductance via the
Thouless formula. Indeed, the larger the level spacing the greater the
promotional energy cost, and hence the smaller the capacitance. Concomitantly, the
larger energy spacing implies larger resistance. The two are functionally
related, and conspire to give the RC-time constant that scales with the
sample size as derived above for the three metallic regimes.

As for the observability of the quantum capacitance, we note that it
combines with the classical (geometrical)  capacitance in series.$^5$
Thus, it is effective only for sufficiently small samples inasmuch as
\(C_Q = e^2/2\pi\Delta \sim L^3\). while \(C_{classical} \sim L\) for a
three-dimensional mesoscopic sample. 

Finally, our present approach to mesoscopic quantum capacitance, based as
it is on the Thouless conductance formula, shares the latter's generality
inasmuch as the single electron coupling and the level spacing may be
generalized to the coupling of correlated electron excitations and the
associated excitation energy-spacings.$^8$

In conclusion, we have derived the statistics of mesoscopic capacitance
fluctuations basing directly  on the generally valid Thouless conductance
formula by identifying the escape time as the {\sl RC-time constant}.
Inasmuch as this quantum capacitance combines in series with the classical
geometrical capacitance, the speed of a mesoscopic circuit element must
be limited ultimately by this quantum {\sl RC-time constant}.

\end{document}